# Dynamics of heterogeneous hard spheres in a file

Ophir Flomenbom

*Flomenbom-BPS, 19 Louis Marshal St., Tel Aviv, Israel, 62668*

**Abstract** – Normal dynamics in a quasi-one-dimensional channel of length $L$ ($\to\infty$) of $N$ hard spheres are analyzed. The spheres are heterogeneous: each has a diffusion coefficient $D$ that is drawn from a probability density function (PDF), $W \sim D^{-\gamma}$, for small $D$, where 0≤$\gamma$<1. The initial spheres' density $\rho$ is non-uniform and scales with the distance (from the origin) $l$ as, $\rho \sim l^{-a}$, 0≤$a$≤1. An approximation for the $N$-particle PDF for this problem is derived. From this solution, scaling law analysis and numerical simulations, we show here that the mean square displacement for a particle in such a system obeys, <$r^2$>~$t^{(1-\gamma)/(2c-\gamma)}$, where $c = 1/(1 + a)$. The PDF of the tagged particle is Gaussian in position. Generalizations of these results are considered.

I. INTRODUCTION

Diffusion is among the fundamental processes in condensed matter physics, chemistry and biology, as it affects the behavior of many complex processes in these fields, e.g. [1-4]. An important process in the study of diffusion is file dynamics (also known as single file dynamics) [4-38, 40]. Put simply, it is a process of *N* identical particles (hard spheres) that perform normal stochastic diffusion, with the same diffusion coefficient *D*, in a cylinder, or a strait, of length *L* ($L \to \infty$). The mean particles' density, $\rho$, is fixed: $\rho=\rho_0=N/L$. (This means that the mean microscopic distance between adjacent hard spheres is fixed and follows, $\Delta=L/N$, where $\Delta$ can't be smaller than the particle's diameter). The dynamics of hard spheres in a strait is a very realistic model for many microscopic processes [1, 30-37]; for example: *(a)* diffusion within biological and synthetic pores, and in porous materials, of water, ions, proteins, and organic molecules [1, 30]. (*b*) Diffusion along 1D objects, such as the motion of motor-proteins along filaments [1]. (*c*) Conductance of electrons in nano-wires [37]. *(d)* Single file dynamics has also been related to monomer dynamics in a polymer: both systems share a similar scaling law for the MSD of a tagged monomer [29, 34].

The most well-know property of file-dynamics is the scaling of the mean square displacement (MSD) $<r^2>$ of a tagged particle in the file: $<r^2> \approx (Dt)^{1/2}/\rho_0$. This result is unique. It is much slower than the MSD of a free meso-scopic particle diffusing in solution, for which, $<r^2>_{free} \approx Dt$. Clearly, a tagged particle in a file is much slower than a free particle as it can only move when other particles move in the same direction. Still, the special scaling of $<r^2>$ with time reflects a unique mechanism of motion. In Ref. [23], we have derived a general relation connecting the mean absolute displacement (MAD) of a free particle and of a tagged particle in a file (that have the same underlying dynamics) that captures some of this uniqueness:

$$<|r|> \approx <|r|>_{free}/n. \qquad (1)$$

Here $n$ is the number of particles in the covered length $<|r|>$. Equation (1) holds when the file has a fixed density on average ($<|r|> \approx n/\rho_0$), and this leads to,

$$<|r|> \approx \rho_0^{-1/2}<|r|>_{free}^{1/2}. \qquad (2)$$

Equations (1)-(2) show that when diffusing a distance $r$, the tagged particle slows down relative to a free particle as it can only move when coordination with the file particles is achieved, and this coordination is proportional to one over the number of particles in the distance $r$. The relation in Eq. (2) leads to the famous MSD in a Brownian file, that is, $<r^2> \approx (Dt)^{1/2}/\rho_0$.

Yet, there are many other known statistical properties of file-dynamics [4-27]: *(a)* The PDF of the tagged particle is asymptotically a Gaussian in position [5]. *(b)* The motion of the particles is correlative, namely, a cloudlike-motion is seen in the system [9, 18]. This cloud of particles is not of a constant density, namely, fluctuations in the particles' density are observed [9, 18]. *(c)* The microscopic single event PDFs in time and space have finite moments [17]. *(d)* In dimensions larger than one, a tagged hard sphere in the presence of hard spheres diffuses normally [9]; namely, in such a system the MSD of a tagged particle is linear with time. *(e)* For a deterministic basic single file with momentum exchange upon collisions, the tagged particle's PDF is also a Gaussian, yet with a variance that scales as the time [6] (note that Eq. (2) still holds). *(f)* We note that in this Paper, the statistics of the particles at the edges of the file are not considered as special particles. Indeed, in a file with a finite number of particles, yet of infinite length, the particles at the edge of the file can diffuse freely to the side not bounded by particles. For an analysis that focuses on this point, see Ref. [25]. Here, we focus on files with $N \to \infty$. In this regard, the tagged particle represents the particles in the middle of the file.

Still, in realistic systems, one, or several, of the conditions defining the *basic* file may break down, and this may lead to different dynamical behaviors. For example, in a real channel, the particles may bypass each other with a constant probability upon collisions [19-22], and this leads to an enhanced diffusion. Yet, when the particles interact with the channel, a slower diffusion is seen [15]. An important generalization in file-dynamics takes the initial particles' density law to scale with the distance $l$ [23],

$$\rho(l) = \rho_0 (l/\Delta)^{-a} \quad ; \quad 0 \leq a \leq 1. \tag{3}$$

$\rho(l)$ in Eq. (3) is the initial density of the file: the particles are initially positioned at, $x_{0,j}=\text{sign}(j)\Delta|j|^{1/(1-a)}$, for $|j| \leq M$, $N=2M+1$. So, the initial number of particles $n$ as a function of the length $l$ obeys, $n = (l/\Delta)^{1-a}$. Among the possible realistic choices for a particle-distance law (e.g. an exponential, a Gaussian, or a power-law), the one that affects the dynamics is a power-law. This is shown when calculating the MAD for a system obeying Eq. (3) [23],

$$<|r|> \approx \rho_0^{(a-1)/2} <|r|>_{free}^{(1+a)/2}. \tag{4}$$

When $a \to 0$, we recover the standard result, $<|r|> \approx \rho_0^{-1/2} <|r|>_{free}^{1/2}$. This equation means that only a power-law density law can influence the dynamics; namely, when the distance between

particles along the file doesn't increase fast enough, as in a power-law density law, the scaling of the MAD is not affected by the fluctuations in the distance among particles.

Now, $<|r|>$ in Eq. (4) holds for *any* renewal *N*-body underlying dynamics and for the density in Eq. (3). Here, a renewal file is a file in which all the particles attempt to jump at the same time.

Equation (4) generalizes Eq. (2). Still, this generalization is limited to the other conditions of a basic file. In this Letter, we deal with heterogeneous files. In a heterogeneous file, the particles' diffusion coefficients are distributed according to a PDF; here, we use,

$$W(D) = \frac{1-\gamma}{\Lambda}\left(\frac{D}{\Lambda}\right)^{-\gamma}, \quad 0 \leq \gamma < 1, \tag{5}$$

where $\Lambda$ is the fastest possible diffusion coefficient in the file. The initial conditions are distributed according to Eq. (3). In a series of analytical and numerical calculations, we show here that the MSD for the tagged particle in such a file follows,

$$\rho_0^2 <r^2> \sim (\rho_0^2 \Lambda t)^{(1-\gamma)/(2c-\gamma)} \quad ; \quad c = 1/(1+a). \tag{6}$$

The corresponding PDF is a Gaussian. Generalizations and implications of these results are considered.

## II. CALCULATING THE FILE'S PDFS

In this paragraph we calculate the PDF of the tagged particle in a heterogeneous file from the joint PDF for all the particles in the file, $P(x, t \mid x_0)$. Here, $x = \{x_{-M}, x_{-M+1}, \ldots, x_M\}$ is the set of particles' positions at time $t$, and $x_0$ is the set of the particles' initial positions at the initial time, $t_0$, which is set to zero. The tagged particle is taken as the middle particle in the file. The following calculations for $P(x, t \mid x_0)$ are based on our analysis of simple files [23], and so we concisely present these calculations first; the curious reader can also find an elaborated discussion of our previous calculations in Appendix A of the EPAPS Document accompanied this paper [42].

**Simple files.-** In a simple file, $P(x, t \mid x_0)$ obeys a simple normal diffusion equation,

$$\partial_t P(x, t \mid x_0) = D \sum_{j=-M}^{M} \partial_{x_j} \partial_{x_j} P(x, t \mid x_0). \tag{7}$$

Equation (7) is solved with the appropriate boundary conditions, which reflect the hard-sphere-nature of the system: $(D \partial_{x_j} P(x, t \mid x_0))_{x_j = x_{j+1}} = (D \partial_{x_{j+1}} P(x, t \mid x_0))_{x_{j+1} = x_j}$ for $j = -M, \ldots, M-1$, and with the appropriate initial condition:

$$P(x, t \to 0 \mid x_0) = \prod_{j=-M}^{M} \delta(x_j - x_{0,j}). \tag{8}$$

The PDFs' coordinates must obey the order: $x_{-M} \leq x_{-M+1} \leq \cdots \leq x_M$. The solution of Eq. (7) is a sum of products of Gaussians [23-27],

$$P(x, t \mid x_0) = \frac{1}{c_N} \sum_p e^{\frac{-1}{4Dt} \sum_{j=-M}^{M} (x_j - x_{0,j}(p))^2}. \tag{9}$$

In Eq. (9), the external sum is over $N!$ permutations of the initial conditions. The factor that takes care for the normalization is $c_N$; $c_N$ is always the normalization everywhere it appears in this paper. Equation (9) is understood under the condition that the coordinates are ordered. Equation (9) is a direct result of the Bethe ansatz for linearly coupled particles [38].

Equation (9) is the starting point for finding the PDF of a tagged particle in this file, $P(r, t \mid r_0)$. In Ref. [23], we have estimated this PDF as,

$$P(r,t \mid r_0) \approx \frac{1}{c_N} \sum_{\tilde{p}} e^{\frac{-1}{4Dt} \sum_{j=-n}^{n}(r_d - x_{0,j}(\tilde{p}))^2} \leq \frac{1}{c_N} e^{\frac{-r_d^2}{2Dt} \sum_{j=1}^{n} 1}. \tag{10}$$

In Eq. (10), $r_d = r - r_0$. Equation (10) is a result of lengthy calculations, and assumes the limit of long times. The full details of the calculations that relate Eq. (9) and Eq. (10) were presented in Ref. [23]; yet, these are presented in Appendix A in the EPAPS Document accompanied this paper [42]. In what follows, we highlight the important steps of these calculations. We start with Eq. (9), and first integrate the file-coordinates excluding the tagged particle's coordinate. Then, we count the important permutations that contribute to the sum of permutations, after the integration; these then form the values of $\tilde{p}$. Once we know the set $\{\tilde{p}\}$, we can further estimate $P(r, t \mid r_0)$ with the inequality. The inequality simplifies the expression for $P(r, t \mid r_0)$, as the last term in Eq. (10) is a summation over a constant; namely, the sum counts particles, and so its solution is $n$: the number of particles in the length $\tilde{r}$. $\tilde{r}$ is found from the equation,

$$\frac{\tilde{r}(n)}{\sqrt{4Dt}} = 1.$$

This relation for $\tilde{r}$ is a result of our approximation that each exponential factor is a kind of a step function, where the step function is non-zero for a width equal to the variance of the exponential argument. As in a constant density file the distance is proportional to the number of particles in it, $n \sim \tilde{r}/\Delta$, we have, $n \sim \sqrt{Dt}/\Delta$, and thus,

$$P(r,t \mid r_0) \leq \frac{1}{c_N} e^{\frac{-r_d^2}{2Dt} n} = \frac{1}{c_N} e^{\frac{-R_d^2}{\sqrt{2\tau}}},$$

where, $R_d = r_d/\Delta$ and $\tau = \Delta^{-2}Dt$ are the dimensionless distance and time respectively.

***Heterogeneous files.-*** Once the relation connecting Eq. (9) and Eq. (10) is established, we can use a corresponding relation for deriving the PDF of the tagged particle in a heterogeneous file. Clearly, we need first to solve the equation of motion for the *N*-particle PDF for this file:

$$\partial_t P(\boldsymbol{x}, t \mid \boldsymbol{x_0}) = \sum_{j=-M}^{M} D_j \partial_{x_j} \partial_{x_j} P(\boldsymbol{x}, t \mid \boldsymbol{x_0}), \tag{11}$$

subjected to the boundary conditions:

$$(D_j \partial_{x_j} P(\boldsymbol{x}, t \mid \boldsymbol{x_0}))_{x_j = x_{j+1}} = (D_{j+1} \partial_{x_{j+1}} P(\boldsymbol{x}, t \mid \boldsymbol{x_0}))_{x_{j+1} = x_j} \quad ; \quad j = -M, \ldots, M-1, \tag{12}$$

and with the initial condition, Eq. (8). We *approximate* the solution of Eqs. (11)-(12) with,

$$P(\boldsymbol{x}, t \mid \boldsymbol{x_0}) \approx \frac{1}{c_N} \sum_p e^{-\sum_{j=-M}^{M} \frac{(x_j - x_{0,j}(p))^2}{4tD_j}}. \tag{13}$$

Equation (13) is our first main result in this paper. This equation was written in analogy to Eq. (10). To test the quality of the approximation, we plug it in the diffusion equation for a heterogeneous file, Eq. (11). We find that Eq. (13) indeed fulfills Eq. (11). Equation (13) also fulfills the initial condition, Eq. (8). Yet, Eq. (13) only approximates the boundary conditions, Eq. (12). Nevertheless, a simple analysis shows that the approximation in Eq. (13) becomes more and more accurate for large times. (A full analysis of Eq. (13) is presented in appendix B in the EPAPS Document accompanied this paper [42].)

Using Eq. (13), we approximate the PDF of the tagged particle in the heterogeneous file with,

$$P(r, t \mid r_0) \approx \frac{1}{c_N} \sum_{\tilde{p}} e^{-\sum_{j=-n}^{n} \frac{(r_d - x_{0,j}(\tilde{p}))^2}{4tD_j}} \leq \frac{1}{c_N} e^{\frac{-R_d^2}{4\tau} \sum_{j=1}^{n} 1/D_j}. \tag{14}$$

Here, $\tau = \Delta^{-2} \Lambda t$. Equation (14) is based on the same approach that relates Eq. (9) to Eq. (10). (Additional technical comments on this relation are presented in Appendix C in the EPAPS Document accompanied this paper [42].) Yet to proceed, we need to calculate the sum in last factor in Eq. (14). These calculations are more complicated than those performed for the simple file. Firstly, for a heterogeneous file that its diffusion coefficients are drawn from Eq. (5), any group of $n$ particles (taken from the $N$ particles in the file) must have the following values for their diffusion coefficients,

$$D_j \approx \Lambda(1 - (j-1)/n)^{1/(1-\gamma)} \quad ; \quad 1 \leq j \leq n,$$

where the values of the diffusion coefficients are ordered from the largest to the smallest. This relation's accuracy increases as $n \to \infty$. (For the derivation, see Appendix D in the EPAPS Document accompanied this paper [42]), Secondly, we need to find $n(t)$. This is found from the equation:

$$\frac{\tilde{r}(n)^2}{\widetilde{D}_n} = t. \tag{15}$$

Relation (15) represents the arguments in all the exponentials in Eq. (14). $\tilde{r}(n)$ is simply found from the density law in the system, $n \approx (\tilde{r}/\Delta)^{1-a}$. The diffusion coefficient $\widetilde{D}_n$ appearing in Eq. (15) must represent a bunch of slow particles in the interval that has in it $n$ particles, as these particles affect the result the most. Yet, $\widetilde{D}_n$ is a typical slow diffusion coefficient, and not necessarily the slowest. We estimate $\widetilde{D}_n$ as, $\widetilde{D}_n = \Lambda n^{-\gamma/(1-\gamma)}$. The derivation of this relation is spelled out in the next paragraph (*Scaling law analysis*). Here we note that as $\gamma$ tends to one, $\widetilde{D}_n$ reaches the value of the slowest diffusion coefficient from a group of $n$ particles. Yet, for a relative fast system $\widetilde{D}_n$ approaches a constant independent of $n$. A similar trend in seen in the behavior of the average diffusion coefficient, which vanishes as $\gamma$ goes to one and has a non-zero value as $\gamma$ tends to zero. Now, using the above expressions for $\tilde{r}(n)$ and $\tilde{D}_n$ in Eq. (15), we find,

$$n \approx \tau^{\frac{(1-a)(1-\gamma)}{2-\gamma(1+a)}}. \tag{16}$$

Substituting Eq. (16) in Eq. (14) yields the PDF for the tagged particle in a heterogeneous file:

$$P(r,t \mid r_0) \leq \frac{1}{c_N} e^{\frac{-R_d^2}{4\tau} \sum_{j=1}^{n}(1-\frac{j-1}{n})^{\frac{-1}{1-\gamma}}} = \frac{1}{c_N} e^{\frac{-R_d^2}{4\tau} n^{\frac{1}{1-\gamma}}} = \frac{1}{c_N} e^{\frac{-R_d^2}{4\tau} \tau^{\frac{(1-a)}{2-\gamma(1+a)}}}. \tag{17}$$

A Gaussian PDF is specified through its variance, and so,

$$\langle R_d^2 \rangle = 2\tau^{\frac{1-\gamma}{2c-\gamma}}, \qquad c = 1/(1+a). \tag{18}$$

Equations (17) and (18), together with Eq. (13), are the major results in this paper. Note that Eq. (18) is obtained from Eq. (17), and so it is the upper bound of the MSD of this file. Yet, we show in what follows, in scaling law analysis and in simulations, that this is in fact the asymptotic limit of the actual MSD.

Examining Eq. (18), we note the following. In the limit of, $\gamma \to 0$, $<R_d^2> \sim \tau^{(1+a)/2}$. This result is equivalent to Eq. (4) for a Brownian file. This result means that when there are not enough slow particles in the file, the MSD scales in the same way as of a simple file. Thus, this result gives the criteria when $W(D)$ affects the diffusion process significantly. In the limit of a constant density, $a=0$, we have, $<R_d^2> \approx \tau^{(1-\gamma)/(2-\gamma)}$. Here, when, $\gamma \to 1$, $<R_d^2> \approx 1$, namely, in this limit the system is frozen. Equation (18) also predicts a cancellation of opposing effects: slow diffusion due to many slow particles and fast diffusion due to a low particles' density can cancel each other;

when: $a = \gamma/(2 - \gamma)$, a simple file scaling is seen, $<R_d^2> \sim \tau^{1/2}$, yet the actual file is heterogeneous.

Finally, we note here that a very different result for the MSD than Eq. (18) is obtained in a heterogeneous file obeying Eq. (5), when all the particles start at the origin; see Ref. [25] for a discussion.

### III. SCALING LAW ANALYSIS

In this paragraph, we derive a scaling law for $<|r|>$ in a heterogeneous file with a constant density. The results of this paragraph support Eq. (18), and further illuminate heterogeneous files. We start with the following set of relations,

$$<|r|> = <|r|>_{free}/n = \Delta^{1/2}<|r|>^{1/2}_{free} \approx \Delta^{1/2}[D(<|r|>_{free})t]^{1/4}. \tag{19}$$

Equation (19) is similar to Eq. (1): $n$ is the number of particles in the cover length, yet $<|r|>_{free}$ reflects a free particle dynamics with a *modified diffusion coefficient*, $<|r|>_{free} \approx [D(<|r|>_{free})t]^{1/2}$. $D(<|r|>_{free})$ should reflect the fact that in an interval of length $<|r|>_{free}$ there is a typical diffusion coefficient that represents all the particles in this length, as we substitute one for many. Clearly, $D(<|r|>_{free})$ is among the slowest ones in the interval $<|r|>_{free}$. Still, it should represent a bunch of slow particles, and not merely the slowest one. To estimate $D(<|r|>_{free})$, we first derive the PDF of the smallest diffusion constant, $D_{min}$, among $n$ particles, denoted as $f(D_{min}, n)$. The diffusion coefficients of the particles are drawn independently of each other, and so this PDF obeys,

$$f(D_{min}, n+1) = W(D_{min})(\int_{D_{min}}^{\Lambda} W(D)dD)^n. \tag{20}$$

The factor $W(D_{min})$ is the PDF that the slowest diffusion coefficient has a value of $D_{min}$ and the integral to the power of $n$ is the probability that all the other particles have diffusion coefficients that are larger than $D_{min}$. A normalization constant doesn't affect the following calculations, and it is omitted. Using Eq. (5) in Eq. (20), we find (for $n \gg 1$),

$$f(D_{min}, n+1) \approx (D_{min}/\Lambda)^{-\gamma} e^{-n(D_{min}/\Lambda)^{1-\gamma}}. \tag{21}$$

Equation (21) has the typical form of a PDF in extreme value statistics [38]. We use this PDF to link a typical small diffusion coefficient to $n$. For this, we look on the exponential factor in the PDF, $e^{-n(D_{min}/\Lambda)^{1-\gamma}}$, and notice that only when the condition, $n(\widetilde{D}_{min}/\Lambda)^{1-\gamma} = 1$, is met, a large probability can be assigned for small values of $D_{min}$. Solving for $\widetilde{D}_{min}$, we find, $\widetilde{D}_{min} = \Lambda n^{-1/(1-\gamma)}$. Using $\widetilde{D}_{min}$ in Eq. (21) leads to,

$$f(\widetilde{D}_{min}, n) \approx \Lambda^{-1} n^{\gamma/(1-\gamma)}. \tag{22}$$

We define the typical value for the slowest particles in the interval of $n$ particles ($n \gg 1$), denoted as $\widetilde{D}_n$, as one over the PDF $f(\widetilde{D}_{min}, n)$,

$$\widetilde{D}_n \equiv 1/f(\widetilde{D}_{min}, n) \approx \Lambda n^{-\gamma/(1-\gamma)}. \tag{23}$$

Equation (23) was used in the previous paragraph to derive Eq. (17). Substituting Eq. (23) into Eq. (19), with $D(\langle |r| \rangle_{free}) \to \widetilde{D}_n$ and $n$ in Eq. (16), leads to,

$$\langle |R_d| \rangle = \tau^{\frac{1-\gamma}{2(2-\gamma)}}. \tag{24}$$

Equation (24) is the same as Eq. (18) for $a = 0$, with, $\langle R_d^2 \rangle \approx \langle |R_d| \rangle^2$. Namely, Eq. (24) supports the results obtained in the previous paragraph. Indeed, both calculations rely on the same form

for $\tilde{D}_n$, yet these calculations have different starting points. Note that the scaling law considered here holds for a=0. In a file with a non-uniform particles' density, the file's density doesn't scale with the distance in the sense that a given interval of length $l$ taken from the file at different locations along the file has a different density of particles. Thus, any scaling law for a non-fixed density file must rely significantly on known results. Starting from Eq. (19), we do not need to rely on known results. Yet, the reader can find in Ref. [23] a scaling law analysis that uses also known results, in deriving scaling laws for non-uniform files.

Scaling law analysis enables to generalize the results for files with different kinds of dynamics. We consider in what follows heterogeneous-deterministic files. A deterministic file is a file in which the particles are Newtonian and each particle is assigned an initial velocity $\pm v$ with equal probability. In a simple deterministic file, the PDF of a tagged particle is a Gaussian with a variance that scales linearly with time. What is $\langle |R_d| \rangle$ when the value $|v|$ is drawn from a PDF of the form of Eq. (5) with equal probability for any direction? Starting from Eq. (19), we find,

$$\langle |R_d| \rangle = (\Delta^{-1} |\tilde{v}| t)^{\frac{1-\gamma}{2-\gamma}}, \qquad (25)$$

where $|\tilde{v}|$ is a characteristic velocity in the system. Equation (25) is calculated in a similar way to the analysis of this paragraph. Equation (25) shows that as γ→1 the deterministic file freezes and as γ→0 the file behaves as a simple deterministic file.

**IV. NUMERICAL SIMULATIONS**

We perform *off*-lattice simulations of Eq. (11) with hard core interactions between point particles. The fact that the particles are point-like reflects the equation of motion, yet, does not change the long time statistics of the file compared to simulations of files on lattices. (In fact, simulations are always lattice-like as the smallest length scale is limited by the precision of the machine.) In the simulations, each particle is assigned a diffusion coefficient from the PDF in Eq. (5) ($\Lambda$=1 in the simulation). The $j^{th}$ particle is positioned at, $x_{0,j}$=sign(j)|j|$^{1/(1-a)}\Delta$ ($\Delta$=1.3 in the simulation). We set *N=501* particles. In each time step (*dt* = 0.13 in the simulations), each particle is moved relative to its position according to the equation, $dx_j = 2(q - 1/2)\sqrt{2D_j t}$, where $q$ is a random number from the unit PDF, and is chosen for each particle at each time step. The particles' locations are ordered after each time step. The interval's length is bound: edges particles can't move further than their initial conditions plus a room for several full jumps in the direction that extends the initial interval length. The above iteration scheme is executed over and over and over again (three millions time steps are used in each simulation). Note that in the above simulations' rules, the boundary conditions are *always* fulfilled. Also, note that the above simulations' rules were also used for simple files; e.g. files with the same diffusion coefficient. Yet, these rules hold also for the heterogeneous file. Here, the reflection principle (that is, the ordering of the particles after each cycle of jumps) represents: (a) elastic collisions among particles that can clearly also represent particles with distribution of diffusion coefficients, and (b) Brownian dynamics, so the particles momenta decay after each jump relatively fast, and so in the next cycle of jumps the particles do not drag previous velocities.

We perform extensive simulations. Each simulation has different values for *a* and *γ* where, *a=0, 1/3, 2/3,* and, *γ=0, 1/3, 1/2, 2/3*. In each simulation, we calculate the MSD for thirty particles from the file. For each simulation (defined with a specific values for *γ* and *a*), the run time for the simulation and the MSD-calculations is three minutes on a standard 3.33 GHz PC.

Figure 1 presents the results for the MSD from all the simulations. Each panel shows MSD-curves for three values of *a* each with the same value of *γ*. The analytical curves obtained from Eq. (18) are also shown. The curves coincide with the numerical results to a satisfactory level. The only point to note is that as *a* increases, converges occurs at larger times. This is an expected behavior for a file with non-fixed particle's density.

In light of the simulations' results, a final remark is made on the interpretation of the limit of long times. In this paper, we used this limit in deriving the statistics of the file. We gave along the paper, and in the appendices, several interpretations for this limit. Yet, we can use Fig. 1 for further define the meaning of long times. Figure 1 shows that this limit depends on the value of $\gamma$ and $a$: when $\gamma$ and/or $a$ are large, the coincidence of the simulations' curves and the curves obtained from Eq. (18) happens at relatively larger times; plus, at smaller times, the difference among the curves is, in most cases, larger when $\gamma$ and $a$ are larger. So, we say that long times corresponds to the time, $t^*$, it takes a particle to reach a distance $r^*$ from its origin that has $n^*$ particles in it. $t^*$ is then estimated with Eq. (18): $r^* \sim \Delta n^{*1/(1-a)} \sim \Delta\sqrt{\langle R_a^2 \rangle}$. These relations give: $t^* \sim \sim \frac{\Delta^2}{\Lambda} n^{*2/[\mu(1-a)]}$, where $\mu$ is the scaling power in Eq. (18). We use $n^* = 35$ as a safe-bound for $t^*$, as the value of 35 (events) is considered a large number in statistics. From Fig. 1, it is clear that also for $n^* \approx 9$ the coincidence among the simulations' results and the curves obtained from Eq. (18) is excellent.

## V. CONCLUDING REMARKS

This paper deals with normal stochastic dynamics of heterogeneous hard spheres in a very long strait. Each sphere has a random diffusion coefficient drawn from a PDF, $W(D) \sim D^{-\gamma}$, $0 \leq \gamma < 1$, for small $D$. The initial positions are also distributed such that the initial particles' density law obeys, $\rho(l) \sim \rho_0 (l/\Delta)^{-a}$, $0 \leq a \leq 1$, where $l$ is the distance from the orgin. We first derive the approximation for the particles' PDF for heterogeneous files: $P(x, t \mid x_0) \approx \frac{1}{c_N} \sum_p e^{\frac{-1}{4t} \sum_{j=-M}^{M} \frac{(x_j - x_{0,j}(p))^2}{D_j}}$. From this PDF, we derive the statistics of the tagged particle in heterogeneous files: $\langle R_d^2 \rangle = 2\tau^{\frac{1-\gamma}{2c-\gamma}}$, $c = 1/(1+a)$, and, $P(r, t \mid r_0) \sim \frac{1}{c_N} e^{\frac{-R_d^2}{2 \langle R_d^2 \rangle}}$. The same results for the tagged particle's MSD were obtained using additional two approaches: scaling law analysis and numerical simulations. We also obtained results for deterministic files with a constant particles' density and distribution in velocities of the form of Eq. (5); here, using scaling law analysis, we found that the MAD obeys: $\langle |R_d| \rangle \sim \tau^{\frac{1-\gamma}{2-\gamma}}$. All the above results are useful for files in which the particles are not identical, and differ in, for example, mass, size, or composition.

Still, there is an interesting generalization of the above: anomalous files. In an anomalous file, the underlying dynamics are such that the waiting time PDF for individual jumps decays like a power-law. (A waiting time PDF in a Brownian file decays exponentially.) Anomalous files may exhibit a rich spectrum of behaviors. We find in preliminary calculations that the nature of the anomaly of the file determines its statistical behavior. For example, renewal-anomalous files, in which all the particles attempt jumping at the same time, are different than non-renewal anomalous files, where each particle has its own clock of waiting times [28]. Also, anomalous files with fluctuating diffusion coefficients may lead to interesting phenomena; this statement relied on a corresponding system with a free particle: when a free stochastic particle performs anomalous dynamics and its diffusion coefficient is drawn every jump from a distribution, a transition in the rule for the power that governs the effective waiting-time PDF of the dynamics is seen [41]. Further analysis of anomalous files is still to come.

Funding for this work was partially come from The Ministry of Immigrant Absorption of the State of Israel, The Center for Absorption in Science.

## FIGURE CAPTIONS

**Fig 1** (color online) The MSD on a log-log scale from twelve different simulations. Each simulation has a specific value for $a$ and $\gamma$, where: $\gamma$=0, 1/3, 1/2, 2/3, and $a$=0, 1/3, 2/3. Each panel has a constant value of $\gamma$ (the smallest value of $\gamma$ is in the top-right panel and $\gamma$ increases in a z-like shape). Each curve (in a given panel) corresponds to a different value of $a$, where a lower curve always has a smaller value of $a$. The analytical curves from Eq. (18) are also shown, and coincide nicely with the results from the simulations. [The free parameter of any analytical curve is always chosen to coincide best with the curve of the simulation. Yet, the curve's slope is obtained from Eq. (18).] Note that the *x* axis in the figure was obtained when monitoring the value of $t_j$ every, $10^{Aj}$ time units (*A* is a number), and then taking the log of the time vector. The *Y* axis is the log of the monitored MSD.

# FIGURES

FIGURE1

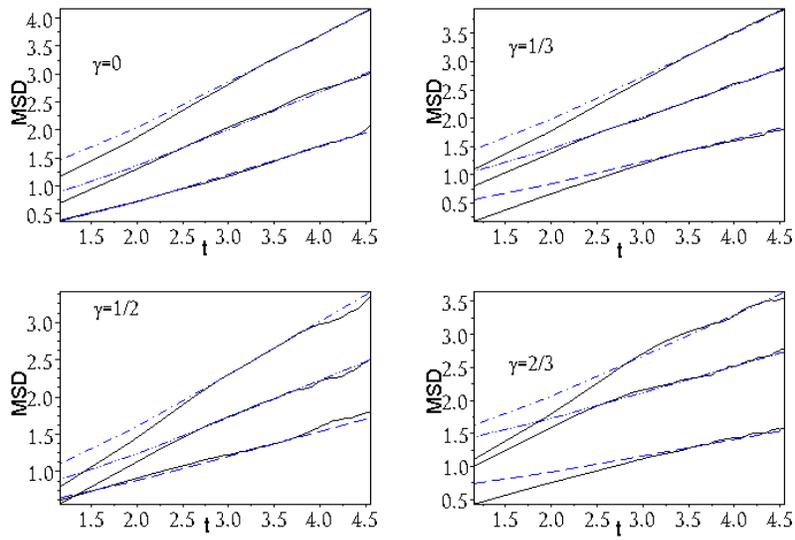

**Supplementary material for the paper:**

**Dynamics of heterogeneous hard spheres in a file**

Ophir Flomenbom

*Flomenbom-BPS, 19 Louis Marshal St., Tel Aviv, Israel, 62668*

**APPENDIX A**

This appendix summarizes the results of our previous paper in this subject, Ref. [1], where we derived the PDF $P(r,t|r_0)$ for a tagged particle in a file of particles all having the same diffusion coefficient. Very similar calculations are used in this paper for deriving PDFs in heterogeneous files; see the next appendix and the calculations in the main text, Eqs. (11)-(18).

The diffusion equation for the particles' PDF $P(\boldsymbol{x},t \mid \boldsymbol{x_0})$ reads:

$$\partial_t P(\boldsymbol{x},t \mid \boldsymbol{x_0}) = D \sum_{j=-M}^{M} \partial_{x_j x_j} P(\boldsymbol{x},t \mid \boldsymbol{x_0}), \tag{A1}$$

with the initial condition,

$$P(\boldsymbol{x},t \to 0 \mid \boldsymbol{x_0}) = \prod_{j=-M}^{M} \delta(x_j - x_{0,j}) \quad ; \quad x_{0,j} = \text{sgn}(j)\Delta|j|^{1/(1-\alpha)}. \tag{A2}$$

Here, $\Delta$ is a microscopic length, and $0 \leq \alpha < 1$. In this appendix, we set, $\alpha = 0$. Each pair of adjacent particles in the file obeys a reflecting boundary condition,

$$[\partial_{x_j} P(\boldsymbol{x},t \mid \boldsymbol{x_0}) - \partial_{x_{j+1}} P(\boldsymbol{x},t \mid \boldsymbol{x_0})]_{x_j = x_{j+1}} = 0 \quad ; \quad -M \leq j < M-1, \tag{A3}$$

which simply means that the adjacent particles bounce back when collide. The joint N-particle multi-dimensional PDF can be obtained from the Bethe ansatz [2]. The Bethe ansatz is the {k}-space, {k}=$k_{-M},…,k_M$, integrand of the Fourier transform of the solution (x→k),

$$\hat{P}(\mathbf{k}, t \mid \mathbf{x_0}) = \frac{1}{N!} \prod_{j=-M}^{M} e^{-ik_j x_{0,j}} e^{-Dtk_j^2} \sum_p e^{ik_j x_j(p)}. \tag{A4}$$

Here, the index $p$ contains the permutations of the N particles' indices, so the summation is over N! permutations (e.g. $x_j(p^*)=x_i$, for a given $p^*$, and, $-M \leq i,j \leq M$).

The joint PDF in {x}-space reads,

$$P(\mathbf{x}, t \mid \mathbf{x_0}) = (4\pi Dt)^{-\frac{N}{2}} \sum_p \prod_j e^{\frac{-\left(x_j - x_{0,j}(p)\right)^2}{4Dt}}. \tag{A5}$$

To show that $P(\mathbf{x}, t \mid \mathbf{x_0})$ is normalized to 1, we need to perform an N-dimensional integration over the {x}-space with the restriction,

$$-\infty \leq x_{-M} \leq x_{-M+1} \leq \cdots \leq x_{M-1} \leq x_M \leq \infty. \tag{A6}$$

It is seen from the direct calculations for small $N$ values that the restricted integration can be replaced with an unrestricted integration for each particle, i.e., $-\infty \leq x_j \leq \infty$, j=-M,…, M, when dividing the result with N!. Thus, each permutation in the expression for $P(\mathbf{x}, t \mid \mathbf{x_0})$ is a product of $N$ integrals, each of which is normalized to one, and so each permutation contributes a factor of 1/N!. As there are N! permutations, $P(\mathbf{x}, t \mid \mathbf{x_0})$ is normalized to one.

For obtaining the PDF for the tagged particle, $P(r, t|r_0)$, $r \equiv x_0$ and $r_0 = 0$, we need to integrate out all the file particles' coordinates except that of $r$, while obeying the above restriction, Eq. (A6). This is performed when separating the integrals into left integrals, and right integrals,

$$P(r,t|r_0) = \int_{-\infty}^{x_{-M+1}} dx_{-M} \int_{-\infty}^{x_{-M+2}} dx_{-M+1} \ldots \int_{-\infty}^{r} dx_{-1} \int_{r}^{\infty} dx_1 \int_{x_1}^{\infty} dx_2 \ldots \int_{x_{M-1}}^{\infty} P(\boldsymbol{x},t \mid \boldsymbol{x_0}) \, dx_M.$$

This $2M$-dimensional integration fulfills Eq. (A6). The particles always maintain their order. Similar with the calculations of the normalization constant, we can use $r$ as the upper bounds in all the left integrals, and use $r$ as the lower bounds in all the right integrals. Then, $P(r,t|r_0)$ obeys,

$$P(r,t|r_0) = \frac{1}{C} \prod_{j=1}^{M} \int_{-\infty}^{r} dx_{-j} \int_{r}^{\infty} dx_j \, P(\boldsymbol{x},t \mid \boldsymbol{x_0}) , \tag{A7}$$

where $C$ is the normalization constant. Equation (A7) enables further analysis because it gives $P(r,t|r_0)$ as products of separate integrals,

$$P(r,t|r_0) \propto \sum_p e^{-\frac{1}{r_f^2}[r-r_0(p)]^2} \prod_{j=1}^{M} \int_{-\infty}^{r} dx_{-j} \, e^{-\frac{1}{r_f^2}[x_{-j}-x_{0,-j}(p)]^2} \int_{r}^{\infty} dx_j \, e^{-\frac{1}{r_f^2}[x_j-x_{0,j}(p)]^2}. \tag{A8}$$

Here, for notation convenience, we define, $r_f \equiv \sqrt{4Dt}$. ($r_f$ equals $<|r|>_{free}$ for normal diffusion, and it is the natural length scale in the system.) For any permutation $p'$, the faith of each integral over $x_j$, with $j > 0$, is one of following three possible outcomes (here we use asymptotic analysis of large times and a finite $r$):

(1) When $(r - x_{0,j})/r_f \to 0$, the integral is approximated with, $\pi/2$.

(2) When $(r - x_{0,j})/r_f \to -\infty$, the integral is approximated with, $\pi$.

(3) When $(r - x_{0,j})/r_f \to +\infty$, the integral is approximated with, $\frac{e^{-Y_j^2}}{2|Y_j|}$, where, $Y_j = (r - x_{0,j})/r_f$.

The same three possible outcomes are obtained for any integral over $x_j$ with $j < 0$, when switching the condition-part of cases (2) and (3). For each permutation, we count the number of integrals of each kind (cases (1)-(3) above), and then sum over all permutations' results. Counting the important permutations that contribute for $P(r,t|r_0)$ in (A8) is the intriguing part in the calculations of this PDF. Yet, once we manage identifying and actually counting these permutations, we can use the same calculations' steps also for heterogeneous files. This is the reason that we spell out these calculations here.

We start the analysis of Eq. (A8) when analyzing $P(r,t|r_0)$ for small values of $r$. Here, small $r$ values are such that, $|r| \leq r_f$. We define *ordered-permutations*, denoted with $\{p_o\}$, as permutations in which all the initial conditions for a positive $j$, $\{x_{0,j}(p_o)\}_{j=1}^{M}$, have positive values, and so they are on the right of $r$, and all the initial conditions for negative $j$, $\{x_{0,-j}(p_o)\}_{j=1}^{M}$, have negative values and are thus located on the left of $r$. Figure 1 of this document illustrates such a possible permutation:

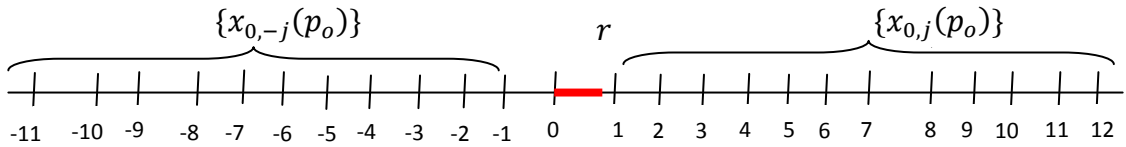

**Fig 1** An illustration of an ordered permutation. Shown are a realization of the initial conditions, with, $x_{0,j}(p_o) = \Delta j$, and the value of $r$ (in red). Each tick represents a particle at the initial stage of the process.

As the tagged particle is the middle particle, and $r$ is small, ordered-permutations exist. In fact, there are $(M!)^2$ such permutations. There are $M!$ internal permutations of the left initial conditions and $M!$ internal permutations of the right initial conditions, starting from the 'perfectly' ordered permutation, $p_o = 1$: $x_{0,j}(1) = \Delta j$ for every $j$. All such $(M!)^2$ permutations of the 'perfectly' ordered permutation lead to the same result of the integrals in Eq. (A8), as the integrals are independent of each other. For small $r$, only cases (1) and (2) are relevant for the ordered permutations. Each ordered permutation gives a constant independent of $r$, which gives, $\left(\frac{1}{2}\right)^{2r_f}$. So, we find that for the ordered-permutations, Eq. (A8) is reduced, and reads:

$$\left(\frac{1}{2}\right)^{2r_f} \sum_p e^{-\frac{1}{r_f^2}[r-r_0(p)]^2} \sim \left(\frac{1}{2}\right)^{2r_f} \frac{1}{C}\int_{-\infty}^{\infty} e^{-\frac{1}{r_f^2}[r-\Delta p]^2} dp = \left(\frac{1}{2}\right)^{2r_f}.$$

That is, the contribution of ordered permutations to Eq. (A8) is a constant independent of $r$.

Thus, for $(M!)^2$ ordered permutations from the possible $(2M)!$ permutations in Eq. (A8), the small $r$ limit contributes a constant. There are still $4^M(M!)^2$ permutations in which the initial conditions are not ordered. For calculating these permutations, we perform the following calculations: we start with the perfectly ordered permutation, and choose $m$ initial coordinates from the left-$M$-initial coordinates, and choose $m$ initial coordinates from the right-$M$-ordered initial coordinates, and switch the sets. For each switch, there are the 'standard' $(M!)^2$ internal permutations all resulting in the same result (that we still need calculating for each switching protocol). We distinguish among the following switching protocols:

- The chosen initial-coordinate is within the distance $r_f$ from $r$: $|r - x_{0,j}(p)| \leq r_f$
- The chosen initial-coordinate is at a distance larger than $r_f$ from $r$: $|r - x_{0,j}(p)| > r_f$

Using these options, we find that there are 4 possibilities for each switch, with the following results:

- The contribution from switching an initial coordinate within the distance of $r_f$ from $r$ with an initial coordinate within the distance of $r_f$ from $r$ from the other side, that is,

$$x_{0,j}(p) \leftrightarrow x_{0,-i}(p) \quad ; r + r_f - x_{0,j}(p) > 0 \text{ and } r - r_f - x_{0,-i}(p) > 0,$$

gives approximately the result of the ordered permutations discussed above, that is, a constant independent of $r$.

- The contribution from permutations in which both initial coordinates that are switched are at a distance larger than $r_f$ from $r$ (in opposite direction), that is,

$$x_{0,j}(p) \leftrightarrow x_{0,-i}(p) \quad ; \quad r + r_f - x_{0,j}(p) < 0 \text{ and } r - r_f - x_{0,-i}(p) < 0,$$

is small relative to the contributions from the switching protocols in (A9) discussed in the next case.

- The important case is when an initial coordinate within the distance of $r_f$ from $r$ is switched with an initial coordinate from the other side (right-left switch or left-right switch) that its distance to $r$ is larger than $r_f$:

$$x_{0,j}(p) \leftrightarrow x_{0,-i}(p) \quad ; r + r_f - x_{0,j}(p) > 0 \text{ and } r - r_f - x_{0,-i}(p) < 0, \quad \text{(A9.1)}$$

or,

$$x_{0,j}(p) \leftrightarrow x_{0,-i}(p) \quad ; r + r_f - x_{0,j}(p) < 0 \text{ and } r - r_f - x_{0,-i}(p) > 0. \quad \text{(A9.2)}$$

In what follows we calculate the contributions from these permutations.

Using the results of case (3) above, we find that the switching protocols (A9.1)-(A9.2) contribute to Eq. (A8) a term that is proportional to the following expression:

$$P(r,t|r_0) \propto (M!)^2 \sum_{z=1}^{n(r_f)} \sum_{q=1}^{n(r_f)-z} S_z S_q \prod_{j=1}^{z} \frac{e^{-[Y_j(p)]^2}}{|Y_j(p)|} \prod_{i=1}^{q} \frac{e^{-[Y_{-j}(p)]^2}}{|Y_{-j}(p)|}. \tag{A10.1}$$

In the upper bounds of the summations in Eq. (A10.1), $n(r_f)$ is the number of particles in $r_f$. In a file with a constant density we have, $\alpha = 0$, and so,

$$n(r_f) = \rho_0 r_f.$$

In Eq. (A10.1), $S_z$ is the combinatorial factor,

$$S_z = \binom{M - \rho r_f}{z}\binom{\rho r_f}{z},$$

which gives the number of ways to perform the switching protocol for $z$ coordinates. Equation (A10.1) has two combinatorial factors: $S_z$ is associated with the switching protocol of Eq. (A9.1) and $S_q$ is associated with the switching protocol of Eq. (A9.2). Each combinatorial factor is associated with a product of Gaussians resulting from calculating the integrals of case (3): $S_z$ is associated with the product, $\prod_{j=1}^{z} e^{-[Y_j(p)]^2}/|Y_j(p)|$, and $S_q$ is associated with the product $\prod_{i=1}^{q} e^{-[Y_{-j}(p)]^2}/|Y_{-j}(p)|$. Note that, in principle, the arguments of the Gaussians depend on the summation index, $Y_j(p) = Y_j[p(z)]$, and $Y_{-j}(p) = Y_{-j}[p(q)]$. Yet, the actual form of $Y_j(p)$, in the context of Eq. (A10.1), should obey,

$$Y_{\pm j}(p) = \frac{r \mp \bar{R}}{r_f},$$

where $\bar{R}$ is a very large number. The reason is that $x_{0,\pm j}(p)$ in, $Y_{\pm j}(p) = (r - x_{0,\pm j}(p))/r_f$, should reflect all the $M$ coordinates from the left (right) of $r$ for $x_{0,j}(p)$ [$x_{0,-j}(p)$], and for this we must use an average quantity, say, $\bar{R}$ ($-\bar{R}$), and this quantity is positive (negative) and large when $M$ is large, since $\bar{R}$ is proportional to $M$. We will use this point in the final step of deriving $P(r,t|r_0)$.

Now, we look on Eq. (A10.1) and note that we can replace $S_z$ and $S_q$ with their maximal value, and write an upper bound for Eq. (A10.1):

$$P(r,t|r_0) \leq \frac{1}{C}\sum_{\tilde{p}} e^{-\sum_{j=-n}^{n} Y_j(\tilde{p})^2 + \ln(|Y_j(p)|)}, \tag{A10.2}$$

where $\tilde{p}$ goes over all the permutations in Eq. (A10.1) (about $n^2$ permutations). Equation (10) in the main text uses Eq. (A10.2) without the logarithmic correction in the exponentials. Equations (A10.1)-(A10.2) are the major results of this appendix.

We further analyze Eqs. (A10) in the limit of many particles, where $M$ is much larger than $r_f$. Then, the symmetric term, $z = q = \rho_0 r_f/2 = \rho_0 \sqrt{Dt}$, dominates the sum in Eq. (A10.1), and we find:

$$P(r,t|r_0) \propto (M! S_{\rho_0\sqrt{Dt}})^2 \prod_{j=1}^{\rho_0\sqrt{Dt}} \frac{e^{-[Y_j(p)]^2-[Y_{-j}(p)]^2}}{|Y_j(p)||Y_{-j}(p)|}$$

$$\lesssim (M! S_{\rho_0\sqrt{Dt}})^2 e^{-\rho_0\sqrt{Dt}[Y_+^2+Y_-^2+\log|Y_-||Y_+|]}. \tag{A11}$$

In Eq. (A11), we used, $Y_{\pm j}(p) \to Y_{\pm} = (r \mp \bar{R})/r_f$, relying on the fact that all the initial coordinates in $\{Y_j\}$ are at a distance of, at least, $r_f$ from $r$, yet the average of all of these is much larger, and proportional to $\pm M$. Thus, the leading term for the PDF of the tagged particle reads,

$$P(r,t|r_0) \propto e^{-\frac{r^2}{\sqrt{4Dt}}}, \tag{A12}$$

with a logarithmic correction in the exponent. For large values of $r$, $|r| \geq r_f$, there are always

$\rho_0 r$ initial coordinate in the left of $r$ (say $r > 0$). This gives rise to a correction term, $e^{-\frac{\rho_0(|r|-\sqrt{4Dt})^3}{4Dt}}$, which multiplies the result of any permutation. But, the switching analysis is the same as discussed above. Thus, the tagged particle's PDF for $|r| \geq r_f$ reads,

$$P(r,t|r_0) \propto e^{-\frac{r^2}{\sqrt{4Dt}} - \frac{\rho_0(|r|-\sqrt{4Dt})^3}{4Dt}}. \tag{A13}$$

The correction term is important only when $|r| \geq 3r_f$, yet the PDF at such distances is of the order of $o(10^{-6})$.

**APPENDIX B**

In this appendix, we show that the PDF,

$$P(\mathbf{x},t \mid \mathbf{x_0}) \approx \frac{1}{c_N} \sum_p e^{-\sum_{j=-M}^{M} \frac{(x_j - x_{0,j}(p))^2}{4tD_j}} \quad ; \quad c_N = (4\pi t)^{N/2} \prod_{i=-M}^{M} D_i, \tag{B1}$$

approximates the PDF for an heterogeneous file in the limit of long times. We also define this limit mathematically.

The actual PDF for the heterogeneous file obeys the diffusion equation (in what follows we use $P(\mathbf{x},t \mid \mathbf{x_0})$ both for the actual PDF and its approximation):

$$\partial_t P(\mathbf{x},t \mid \mathbf{x_0}) = \sum_{j=-M}^{M} D_j \partial_{x_j} \partial_{x_j} P(\mathbf{x},t \mid \mathbf{x_0}). \tag{B2}$$

Equation (B2) is solved with the boundary conditions:

$$(D_j \partial_{x_j} P(\mathbf{x},t \mid \mathbf{x_0}))_{x_j = x_{j+1}} = (D_{j+1} \partial_{x_{j+1}} P(\mathbf{x},t \mid \mathbf{x_0}))_{x_{j+1} = x_j} \quad ; \quad j = -M, \ldots, M-1, \tag{B3}$$

and the initial condition:

$$P(x, t \to 0 \mid x_0) = \prod_{j=-M}^{M} \delta(x_j - x_{0,j}) \quad ; \quad x_{0,j} = \Delta \frac{j}{|j|} |j|^{\frac{1}{1-\alpha}}. \tag{B4}$$

Also, the coordinates in $P(x, t \mid x_0)$ obey the order:

$$-\infty \leq x_{-M} \leq x_{-M+1} \leq \cdots \leq x_{M-1} \leq x_M \leq \infty. \tag{B5}$$

We first show that Eq. (B1) reduces to Eq. (B4) in the limit, $t \to 0$. In this limit we find:

$$P(x, t \to 0 \mid x_0) = \sum_p \prod_{j=-M}^{M} \delta(x_j - x_{0,j}(p)), \tag{B6}$$

since any normalized Gaussian reduces to a Delta function in the limit $t \to 0$:

$$\left( \frac{1}{\sqrt{4\pi D_j t}} e^{-\frac{(x_j - x_{0,j}(p))^2}{4tD_j}} \right)_{t \to 0} \to \delta(x_j - x_{0,j}(p)).$$

Yet, since the solution must always obey Eq. (B5), only the ordered permutation survives, say, permutation $p = 1$, obeying:

$$-\infty \leq x_{0,-M} \leq x_{0,-M+1} \leq \cdots \leq x_{0,M-1} \leq x_{0,M} \leq \infty.$$

Namely, we have:

$$P(x, t \to 0 \mid x_0) = \prod_{j=-M}^{M} \delta(x_j - x_{0,j}(1)),$$

as required.

Next, we show that the PDF in Eq. (B1) fulfills the equation of motion. First, we take its time derivative:

$$I = \partial_t \left( \frac{1}{c_N} \sum_p e^{-\sum_{j=-n}^{n} \frac{(x_j - x_{0,j}(p))^2}{4tD_j}} \right) = \left( \frac{1}{c_N} \right) \sum_p e^{-\sum_{j=-M}^{M} \frac{(x_j - x_{0,j}(p))^2}{4tD_j}}$$

$$+ \frac{1}{c_N} \sum_p e^{-\sum_{j=-M}^{M} \frac{(x_j - x_{0,j}(p))^2}{4tD_j}} \left[ \frac{1}{t} \sum_{i=-M}^{M} \frac{(x_i - x_{0,i}(p))^2}{4tD_i} \right]. \tag{B7}$$

Now, using,

$$\left(\frac{\dot{1}}{c_N}\right) = -\frac{N/2}{t}\frac{1}{\prod_{j=-M}^{M}D_j}\frac{1}{(4\pi t)^{N/2}} = -\frac{N/2}{t}\left(\frac{1}{c_N}\right),$$

we have:

$$I = \frac{1}{c_N}\sum_p e^{-\sum_{j=-M}^{M}\frac{(x_j-x_{0,j}(p))^2}{4tD_j}}\left[\frac{1}{t}\sum_{i=-M}^{M}\left(\frac{(x_i-x_{0,i}(p))^2}{4tD_i}-\frac{1}{2}\right)\right]. \tag{B8}$$

Here, we used the relation, $N = 2M + 1$. Now, applying the operator in the right hand side of Eq. (B2) on Eq. (B1):

$$II = \frac{1}{c_N}\sum_{i=-M}^{M}D_i\partial_{x_i}\partial_{x_i}\sum_p e^{-\sum_{j=-M}^{M}\frac{(x_j-x_{0,j}(p))^2}{4tD_j}},$$

we see,

$$II = \frac{1}{c_N}\sum_{j=-M}^{M}D_j\partial_{x_j}\sum_p -2\frac{x_j-x_{0,j}(p)}{4tD_j}e^{-\sum_{i=-M}^{M}\frac{(x_i-x_{0,i}(p))^2}{4tD_i}}$$

$$= \frac{1}{c_N}\sum_{j=-M}^{M}D_j\sum_p\left(\left(-2\frac{x_j-x_{0,j}(p)}{4tD_j}\right)^2 - \frac{1}{2tD_j}\right)e^{-\sum_{i=-M}^{M}\frac{(x_i-x_{0,i}(p))^2}{4tD_i}}.$$

This equation is rewritten as,

$$II = \frac{1}{c_N}\sum_p e^{-\sum_{i=-M}^{M}\frac{(x_i-x_{0,i}(p))^2}{4tD_i}}\sum_{j=-M}^{M}\left(\frac{(x_j-x_{0,j}(p))^2}{4tD_j}-\frac{1}{2t}\right). \tag{B9}$$

Cleary, Eq. (B9) is the same as Eq. (B8).

Now, for the boundary conditions. Applying on the PDF the left hand side of the boundary condition in Eq. (B3), we find:

$$L = \left(D_i\partial_{x_i}\sum_p e^{-\sum_{j=-M}^{M}\frac{(x_j-x_{0,j}(p))^2}{4tD_j}}\right)_{x_i=x_{i+1}\equiv y} =$$

$$\frac{-2}{t}\sum_p(y-x_{0,i}(p))e^{-\left[\frac{(y-x_{0,i}(p))^2}{4tD_i}+\frac{(y-x_{0,i+1}(p))^2}{4tD_{i+1}}+\sum_{j'=-M}^{M}\frac{(x_j-x_{0,j}(p))^2}{4tD_j}\right]} \equiv \sum_p\tilde{L}(p). \tag{B10}$$

Here, the sum in the exponential is over all the coordinates excluding $x_i$ and $x_{i+1}$. Applying on the PDF the right hand side of the boundary condition, one obtains:

$$R = \left(D_{i+1}\partial_{x_{i+1}} \sum_p e^{-\sum_{j=-M}^{M}\frac{(x_j-x_{0,j}(p))^2}{4tD_j}}\right)_{x_i=x_{i+1}\equiv y} =$$

$$\frac{-2}{t}\sum_p(y-x_{0,i+1}(p))e^{-\left[\frac{(y-x_{0,i}(p))^2}{4tD_i}+\frac{(y-x_{0,i+1}(p))^2}{4tD_{i+1}}+\sum_{j'=-M}^{M}\frac{(x_j-x_{0,j}(p))^2}{4tD_j}\right]} \equiv \sum_p \tilde{R}(p). \tag{B11}$$

Now, we look on permutations $p^\#$ and $p^*$: these are the same excluding the values for $x_{0,i}$ and $x_{0,i+1}$. In particular, for permutation $p^\#$, we set:

$$p^\#: \left\{\{p'\}, x_{0,i}(p^\#) = X, x_{0,i+1}(p^\#) = Y\right\},$$

and for permutation $p^*$, we set:

$$p^*: \left\{\{p'\}, x_{0,i}(p^*) = Y, x_{0,i+1}(p^*) = X\right\}.$$

The set $\{p'\}$ contains permutations of all the initial coordinates excluding those of $x_{0,i}$ and $x_{0,i+1}$. Now, in what follows we divide the summation over $p = 1, \ldots, N!$ permutations to triple summation:

$$\sum_p \tilde{R}(p) = \sum_Y \sum_{X\neq Y} \sum_{p'} \tilde{R}(p', X, Y), \tag{B12.1}$$

and we use the equality,

$$\sum_Y \sum_{X\neq Y} \sum_{p'} \tilde{R}(p', X, Y) = \sum_Y \sum_{X\neq Y} \sum_{p'} \tilde{R}(p', Y, X). \tag{B12.2}$$

(Cleary, the above couple of equations hold also for $\tilde{L}(p)$.) We are going to show that the following relation holds in the limit $t \to \infty$:

$$\tilde{L}(p^{\#}) + \tilde{L}(p^*) \approx \tilde{R}(p^{\#}) + \tilde{R}(p^*). \tag{B13}$$

Proving (B13) is enough for proving that (B1) approximates the boundary conditions (B3), since with the aid Eqs. (B12), Eqs. (B12)-(B13) are the full boundary condition.

Starting from Eq. (B13), we have for the left hand side:

$$\frac{\tilde{L}(p^{\#}) + \tilde{L}(p^*)}{c} = (y - X)e^{-\left[\frac{(y-X)^2}{4tD_i} + \frac{(y-Y)^2}{4tD_{i+1}}\right]} + (y - Y)e^{-\left[\frac{(y-Y)^2}{4tD_i} + \frac{(y-X)^2}{4tD_{i+1}}\right]}, \tag{B14.1}$$

where the right hand side of (B13) reads:

$$\frac{\tilde{R}(p^{\#}) + \tilde{R}(p^*)}{c} = (y - Y)e^{-\left[\frac{(y-X)^2}{4tD_i} + \frac{(y-Y)^2}{4tD_{i+1}}\right]} + (y - X)e^{-\left[\frac{(y-Y)^2}{4tD_i} + \frac{(y-X)^2}{4tD_{i+1}}\right]}. \tag{B14.2}$$

The factor $c$ that appears in Eqs. (B14) reads:

$$c = \frac{-2}{t}\sum_{p'} e^{-\sum_{j'=-M}^{M} \frac{(x_j - x_{0,j}(p'))^2}{4tD_j}}.$$

Now, the exponential factor $e^{-\left[\frac{(y-X)^2}{4tD_i} + \frac{(y-Y)^2}{4tD_{i+1}}\right]}$ goes to unity as $t \to \infty$, and Eq. (B14.1) is rewritten:

$$\left.\frac{\tilde{L}(p^{\#}) + \tilde{L}(p^*)}{c}\right|_{t \to \infty} = (y - X)\left(1 - o\left(\frac{(y-X)^2}{4tD_i} + \frac{(y-Y)^2}{4tD_{i+1}}\right)\right)$$

$$+ (y - Y)\left(1 - o\left(\frac{(y-Y)^2}{4tD_i} + \frac{(y-X)^2}{4tD_{i+1}}\right)\right), \tag{B15.1}$$

and similarly, Eq. (B14.2) reads:

$$\left.\frac{\tilde{R}(p^{\#}) + \tilde{R}(p^*)}{c}\right|_{t \to \infty} = (y - Y)\left(1 - o\left(\frac{(y-X)^2}{4tD_i} + \frac{(y-Y)^2}{4tD_{i+1}}\right)\right)$$

$$+ (y - X)\left(1 - o\left(\frac{(y-Y)^2}{4tD_i} + \frac{(y-X)^2}{4tD_{i+1}}\right)\right), \tag{B15.2}$$

Clearly, Eq. (B15.1) and Eq. (B15.2) are equal in a leading order of $o\left(\frac{1}{t}\right)$, and thus, Eqs. (B15) prove Eq. (B13), and thus prove Eq. (3).

**APPENDIX C**

This appendix relates the PDF for a heterogeneous file,

$$P(\boldsymbol{x}, t \mid \boldsymbol{x_0}) \approx \frac{1}{c_N} \sum_p e^{-\sum_{j=-M}^{M} \frac{(x_j - x_{0,j}(p))^2}{4tD_j}}, \tag{C1}$$

with a PDF of a tagged particle in the file,

$$P(r, t \mid r_0) \approx \frac{1}{c_N} \sum_{\tilde{p}} e^{-\sum_{j=-n}^{n} \frac{(r_d - x_{0,j}(\tilde{p}))^2}{4tD_j}}. \tag{C2}$$

In Eq. (C2), $\tilde{p}$ goes over the relevant permutations (about $n^2$ permutations); see the discussion in appendix A, around Eqs. (A10), for further details on the permutations in $\tilde{p}$.

We also discuss in this appendix the technical details that further approximate $P(r, t \mid r_0)$ with,

$$\frac{1}{c_N} \sum_{\tilde{p}} e^{-\sum_{j=-n}^{n} \frac{(r_d - x_{0,j}(\tilde{p}))^2}{4tD_j}} \leq \frac{1}{c_N} e^{\frac{-R_d^2}{4\tau} \sum_{j=1}^{n} (\Lambda/D_j)}. \tag{C3}$$

In Eqs. (C1)-(C3), $1/c_N$ is always a normalization constant, and in Eq. (C3) $R_d = r_d/\Delta$ (where, $r_d = r - r_0$ is the tagged particle coordinate relative to its initial position) and $\tau = \frac{\Lambda}{\Delta^2} t$ are dimensional distance and dimensional time, respectively. Also, we recall that $\Delta$ ($\equiv 1/\rho_0$) is a microscopic length scale and $\Lambda$ is the fastest diffusion coefficient in the file. The relation connecting Eqs. (C1) and (C3), written in a symbolic way, reads,

$$\frac{1}{c_N}\sum_p e^{-\sum_{j=-M}^{M}\frac{(x_j-x_{0,j}(p))^2}{4tD_j}} \mapsto \frac{1}{c_N}\sum_{\tilde{p}} e^{-\sum_{j=-n}^{n}\frac{(r_d-x_{0,j}(\tilde{p}))^2}{4tD_j}}. \tag{C4}$$

Equation (C4) is the same as the relation connecting the corresponding quantities in a file with the a unique diffusion coefficient; see Appendix A, Eq. (A5) and Eq. (A10.2). In fact, we can carry on precisely the same analysis that was used in obtaining Eq. (A10.1) from Eq. (A5), here, for the heterogeneous file, for deriving Eq. (C6). The fact that $D_j$ appears in the denominator of the exponential does not change the number of the particles in the length $\sqrt{\Lambda t}$, and this is the reason that the same analysis holds for both systems.

Now, for explaining the upper bound of, $\sum_{\tilde{p}} e^{-\sum_{j=-n}^{n}\frac{(r_d-x_{0,j}(\tilde{p}))^2}{4tD_j}}$, in Eq. (C3), we notice that $x_{0,\pm j}(\tilde{p})$ is always at a distance from $r_d$ that is not smaller than $\sqrt{4tD_j}$. In fact, we should set,

$$x_{0,\pm j}(\tilde{p}) \to \pm\bar{R}, \tag{C5}$$

where $\bar{R}$ is a large quantity that is proportional to $M$. The reason is simple: $x_{0,\pm j}(p)$ should reflect all the $M$ coordinates from the left (right) of $r$ for $x_{0,j}(p)$ [$x_{0,-j}(p)$], and for this we must use an average quantity, say, $\bar{R}$ $(-\bar{R})$, and this quantity is positive (negative) and large (in absolute value) when $M$ is large. See also the discussion in appendix A above Eq. (A10.2).

Using Eq. (C5) in the exponentials' arguments in Eq. (C3) gives,

$$(r_d - x_{0,\pm j}(\tilde{p}))^2 = r_d^2 \mp 2r_d\bar{R} + \bar{R}^2, \tag{C6}$$

and so,

$$\frac{1}{c_N}\sum_{\tilde{p}} e^{-\sum_{j=-n}^{n}\frac{(r_d-x_{0,j}(\tilde{p}))^2}{4tD_j}} \mapsto \frac{1}{c_N}e^{-\sum_{j=1}^{n}\frac{r_d^2+\bar{R}^2}{2tD_j}}. \tag{C7}$$

Renormalizing Eq. (C7) with respect to $r_d$ gives Eq. (C3).

**APPENDIX D**

Given the PDF,

$$W(D) = (1-\gamma)\Lambda^{-1}(D/\Lambda)^{-\gamma} \quad ; \quad 0 \le \gamma < 1, \tag{D1}$$

defined in the interval, $0 \le D \le \Lambda$, we draw $n$ random numbers from this PDF. What is the shape of the curve when we plot these random numbers when ordering them from the largest value to the smallest? Answering this question gives $D_j$ that appears just above Eq. (15) in the main text:

$$D_j \approx \Lambda(1-(j-1)/n)^{1/(1-\gamma)}. \tag{D2}$$

This expression's accuracy increases with the value of $n$.

For proving Eq. (D2), we first write the expression for drawing a random diffusion coefficient from $W(D)$ using the unit density,

$$W_u(x) = 1 \quad ; \quad 0 \le x \le 1.$$

We use the relation,

$$W(D)dD = W_u(x)dx,$$

and find,

$$D = \Lambda x_r^{1/(1-\gamma)}. \tag{D3}$$

In Eq. (D3), $x_r$ is a random number drawn from the unit PDF, $W_u(x)$. When there are $n$ random numbers, $\vec{x}_r$ is a vector of length $n$. To proceed, we need to find the functional form of element $j$ in this vector after ordering it from the largest value to the smallest. We call this vector, $\vec{x}_r$ with ordered elements, $\vec{y}_r$. It is clear that the largest value of $\vec{y}_r$ is one; the smallest value is $1/n$ (this is shown in what follows). As the density $W_u(x)$ is fixed, $\vec{y}_r$ must have the form,

$$\vec{y}_r = 1, 1 - \tfrac{1}{n}, 1 - \tfrac{2}{n}, \dots, 1 - \tfrac{n-1}{n}. \tag{D4}$$

Equation (D4) proves Eq. (D2).

For showing that the smallest value in $\vec{y}_r$ is $1/n$, we calculate the PDF of the smallest number from possible $n+1$ random numbers drawn independently from $W_u(x)$:

$$\widetilde{W}_u(x; n+1) = W_u(x) \left( \int_x^1 W_u(y) dy \right)^n \approx W_u(x) e^{-nx}. \tag{D5}$$

Similar with the analysis of extreme value statistics in the main text, we find the typical value of the smallest number drawn from $\widetilde{W}_u(x; n+1)$, $x_{t.s.}$, when first demanding that $e^{-nx}$ is not smaller than $e^{-1}$; namely:

$$x^* \lesssim 1/n.$$

Using this upper bound in the re-normalized PDF in Eq. (D5) gives:

$$\widetilde{W}_u(x^*; n+1) = n. \tag{D6}$$

Finally, the typical smallest value of the vector $\vec{y}_r$, $x_{t.s.}$, is the inverse of $\widetilde{W}_u(x^*; n+1)$ in Eq. (D6), that is,

$$x_{t.s.} = 1/n.$$

It is very simple to see that this analysis is very accurate even for 501 particles (the number of particles used in our simulations) in a simulations. Results from several simulations (for three different values of $\gamma$) are shown in Fig. 2. Coincidence of the simulations with Eq. (D2) is evident.

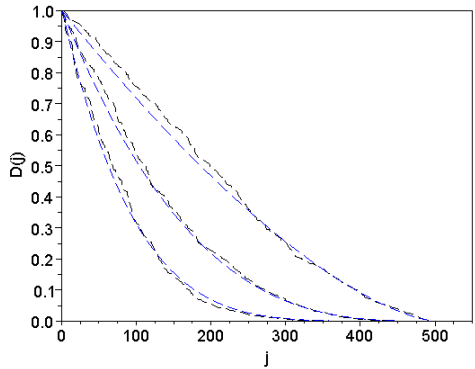

**Fig 2** Results from simulations drawing 501 diffusion coefficients from Eq. (D1), and ordering them from the largest to the smallest; curves are obtained for 3 different values of $\gamma$, $\gamma = \frac{1}{3}, \frac{2}{3}, \frac{2.429}{3}$. The curve with the smaller value of $\gamma$ is to the right of those with lager values of $\gamma$. Here, $\Lambda = 1$. This figure also shows the curves from Eq. (D2) for each value of $\gamma$. Coincidence among the curves from the simulations and the estimated curves is evident.

REFERENCES

[1] O. Flomenbom and A. Taloni, Europhys. Lett. **83**, 20004-p1-p6 (2008).

[2] H. Bethe, Z. Phys. **71**, 205 (1931).